\documentclass[twocolumn,prb]{revtex4}

\usepackage{graphicx}
\usepackage{bm}
\usepackage{amsmath,amssymb,amsfonts}

\renewcommand{\figurename}{\textbf{Figure}}
\renewcommand{\thefigure}{\textbf{\arabic{figure}}}

\renewcommand{\bibnumfmt}[1]{\makebox[1.4em][l]{}} 

\begin{document}

\title{Nonlinear hydrodynamic response of a quantum Hall system}

\author{Hiroki Isobe}
\affiliation{Department of Physics, Kyushu University, Fukuoka 819-0395, Japan}
\affiliation{RIKEN Center for Emergent Matter Science (CEMS), Wako, Saitama 351-0198, Japan}

\begin{abstract}
The quantum Hall effect realizes a quantized Hall resistance $R_{xy} = h/(\nu e^2)$ whereas the longitudinal resistance vanishes.  The quantized value consists of the fundamental physical quantities, the elementary charge $e$ and the Planck constant $h$, along with an integer or fractional constant $\nu$.  
High precision measurements of $R_{xy}$ allude to a linear relation between the applied current $I$ and the Hall voltage $V_\mathrm{H}$.  
Here, we argue that a nonlinear relation between $I$ and $V_\mathrm{H}$ could arise when the electric field is spatially inhomogeneous.  
We first discuss that the linear $I$-$V_\mathrm{H}$ relation holds with Galilean invariance.   
Then we consider a hydrodynamic description of a quantum Hall liquid to deal with an axially symmetric electric field.  It reveals a nonlinear electronic response arising from the centrifugal force exerted on a curved flow and the density gradient invoked by vorticity.  
\end{abstract}

\maketitle

\section*{INTRODUCTION}

The quantum Hall effect realizes a distinct quantum state with a vanishing longitudinal conductance and a quantized Hall conductance. \cite{QHE,FQHE,Ando,Wakabayashi,FQHE2} The accurate quantization of the Hall resistance is used for resistance standard with $R_\mathrm{K} = h/e^2$, consisting of the fundamental constants, the Planck constant $h (= 2\pi\hbar)$ and the elementary charge $e$. \cite{CODATA} Regarding the integer quantum Hall state, Laughlin argued the quantization of the linear Hall conductance by flux insertion \cite{Laughlin}, and it has been realized that a topological invariant known as the Chern number underlays the quantization. \cite{TKNN,Kohmoto} 
For both integer and fractional quantum Hall states, the energy gap of the fermionic many-body ground state protects the quantized charge transport, and hence the quantization is robust against disorder and many-body interaction as long as the bulk energy gap is intact. \cite{Niu-Thouless}
However, such discussions do not prohibit the possibility of nonlinear response in a quantum Hall regime. \cite{He}

The bulk of a quantum Hall system has an energy gap, leading to edge transport  theories. \cite{Halperin,Beenakker,MacDonald} 
Compressible edge states, \cite{Chklovskii} arising from the chiral anomaly, \cite{Kao} can carry an electric current, and the one-dimensional edge transport channels explain the quantized Hall conductance via the Landauer formula. \cite{Streda,Jain-Kivelson,Buttiker}
Nevertheless, the edge transport picture does not exclude macroscopic current transport in the bulk. \cite{Beenakker_review}  
Laughlin's argument \cite{Laughlin} indeed implies Hall current through the incompressible bulk.  
Theories considering the current distribution \cite{MacDonald2,Cage,Tsemekhman} and contactless measurements \cite{Fontein,McCormick,Weis-vonKlitzing} suggest the Hall voltage across the bulk and hence an electric current through the bulk.  

In this work, we demonstrate the nonlinear Hall response in a quantum Hall state by focusing on the bulk contributions to the Hall current.  
We first consider the role of Galilean invariance and revisit Laughlin's argument to examine what imposes the linear Hall response.  Then, we elaborate a hydrodynamic description \cite{Stone,Tokatly2} to elucidate the nonlinear Hall response by illustrating curved Hall current flows.  
Most of our discussions are applicable both to integer and fractional quantum Hall states with bulk energy gaps while the extension of Laughlin's argument is intended for the integer quantum Hall effect.

\section*{RESULTS}

\subsection*{Lorentz force and Galilean invariance} 
We consider a uniform two-dimensional (2D) electron gas confined on the $xy$ plane and apply a uniform magnetic field $\bm{B} = B\hat{z}$ along the $z$ axis.  
We suppose that the electronic system is in a quantum Hall state with the filling factor $\nu$, so that the electron density $\rho$ satisfies $\rho = \nu e B / h$ with the electron charge $e$.  (We use SI throughout.)  
There is no electric field in this inertial frame $\bm{E} = \bm{0}$, and hence the electric current density is $\bm{j} = \bm{0}$.  
We then observe the system in another inertial frame, moving relative to it at velocity $-\bm{v} = -v_x \hat{x}$.  
The Galilean transformation changes the electromagnetic fields to be $\bm{E}' = -\bm{v} \times \bm{B}$ and $\bm{B}' = \bm{B}$.  
Now, electrons move at velocity $+\bm{v}$, and the current density in the moving frame is $j'_x = e\rho v_x = e \rho E'_y / B = (\nu e^2 / h) E'_y$.  Therefore, we find the Hall conductivity $\sigma_{xy} = \sigma_\mathrm{H} = \nu e^2/h$ and the longitudinal conductivity $\sigma_{xx} = 0$.  
The linear relation between the electric field and the Hall current complies with Galilean invariance, provided that the longitudinal conductivity vanishes. \cite{Niu-Thouless,QHE_review}
While we constrain the following discussion to the nonrelativistic regime, we may gain a clearer understanding from a Lorentz transformation \cite{Rosenstein} (Supplemental Section 1).  

\subsection*{Extension of Laughlin's argument} 
We recall Laughlin's argument for the integer quantum Hall state. \cite{Laughlin}  In Figure~1A, a 2D electron gas in an integer quantum Hall state is confined on a Corbino disk.  The system is isolated and conserves the electric charge.   
A magnetic flux $\Phi(t)$ penetrates the hole of the Corbino disk.  As $\Phi(t)$ varies in time, an electric field appears along the azimuthal direction on the disk following Faraday's law of induction: $2\pi r E_\theta = - \dot{\Phi}(t)$ at radius $r$.  
We introduce the polar coordinate $(r,\theta)$ aligned with the disk and the flux points to the $z$ direction.  
The present geometry does not have Galilean invariance, and we thus assume that the Hall current $j_r$ that flows in the radial direction through the bulk obeys a nonlinear relation
\begin{equation}
j_r = \sigma_\mathrm{H} E_\theta + \sigma_2 E_\theta^2 + \sigma_3 E_\theta^3 + \cdots. 
\end{equation}

\begin{figure}
\centering
\includegraphics[width=\hsize]{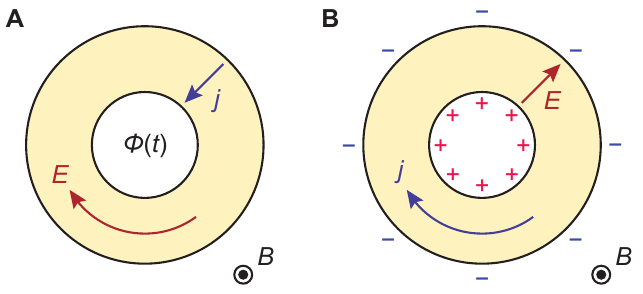}
\caption{%
\textbf{Quantum Hall system in the Corbino geometry}\\
(A) Laughlin's setup to discuss the quantization of the Hall conductance.  The magnetic field $\bm{B}$ points out of the plane.  
\newline
(B) Quantum Hall fluid in the Corbino geometry with the radial electric field.  The Hall current flows in the azimuthal direction. 
}
\label{fig:setup}
\end{figure}

We suppose an adiabatic flux insertion over the time period $0 \leq t \leq T$.  The inserted flux should be a multiple of the flux quantum $h/e$ to assert the gauge invariance, i.e., single-valued electronic wave functions.  
We here set $\Phi(0) = 0$ and $\Phi(T) = h/e$.  
The amount of electric charge $Q$ that passes through a circle of radius $r$ over the time evolution is 
\begin{align}
Q_r &= -2\pi r \int_0^T  j_r \, dt \nonumber\\
&= \int_0^T \left[ \sigma_\mathrm{H} \dot{\Phi} - \frac{\sigma_2}{2\pi r} \dot{\Phi}^2 + \frac{\sigma_3}{(2\pi r)^2} \dot{\Phi}^3 + \cdots \right] dt.  
\end{align}
As the electronic wave function should return to the original one after the flux insertion, the transferred charge should be an integer multiple of the electron charge $Q_r = \nu e$ at \textit{any} $r$, independently of how $\Phi(t)$ evolves.  
It requires $\sigma_\mathrm{H} = \nu e^2/h$ and $\sigma_2 = \sigma_3 = \cdots = 0$.  Therefore, the gauge invariance concludes the exactly linear relation in this setup.  
Instead of the Corbino geometry, a cylindrical geometry as in Laughlin's original work follows the same discussion.

\subsection*{Hydrodynamic analysis} 

Instead of flux insertion that generates an azimuthal electric field, we swap the alignment of the electric field and current and prepare a radial electric field, as shown in Figure~1B.  
Importantly, a flux insertion cannot generate such a electric field, and also the system does not possess Galilean invariance.  The electronic state is not in the equilibrium since an electric current flows under the external electric field.  We assume an axially symmetric electric field $E_r$ and investigate the azimuthal current $j_\theta$ in a quantum Hall state.  
Though the Hall current is dissipationless with the vanishing longitudinal conductivity $\sigma_{xx} = 0$, we retain the electric field to support the steady flow.

We analyze the transport properties from the hydrodynamic viewpoint.  
We suppose that the charged fluid is composed of electrons with electric charge $e$ and consider the momentum transport at scales larger than the magnetic length $l_\mathrm{B} = \sqrt{\hbar/(eB)}$.  
With coarse graining, the local particle density $\rho(\bm{r},t)$ and the velocity $\bm{v}(\bm{r},t)$ describe the fluid motion.  
The latter is tied to the momentum $\bm{p} = m^\ast \bm{v}$, where the effective mass $m^\ast$ serves as the propotionality constant.   
The current density is $\bm{j} = e\rho\bm{v}$.  We constrain the discussion at zero temperature.  

A quantum Hall state is incompressible, and there are two factors that determine the density: the external magnetic field and the vorticity $\omega = \hat{z} \cdot (\nabla \times \bm{v})$.  
They contribute to the density with the units $h/e$ and $h/m^\ast$, respectively. \cite{Stone}  
Therefore, the density of the quantum Hall fluid is 
\begin{equation}
\rho = \frac{eB}{h} + \frac{m^\ast \omega}{h}.
\label{eq:density_constraint}
\end{equation}
For the first term to be dominant, we require $\omega \ll \omega_\mathrm{c}$ with the cyclotron frequency $\omega_\mathrm{c} = eB/m^\ast$.  

As the electron number is conserved locally, the continuity equation $\partial_t \rho + \nabla \cdot (\rho \bm{v}) = 0$ holds.  For an incompressible fluid, the material derivative of the density should vanish: $D\rho/Dt = \partial_t \rho + \bm{v} \cdot \nabla \rho = 0$, leading to a divergence-free flow $\nabla \cdot \bm{v} = 0$.  We note, however, that the density $\rho$ is not necessarily uniform (Supplemental Sections 2 and 4).

Considering the momentum transport in the fluid, we find the Cauchy momentum equation 
\begin{equation}
\rho m^\ast \frac{Dv_i}{Dt} = \partial_j \Sigma_{ij} + f_i, 
\label{eq:EOM}
\end{equation}
where $\Sigma_{ij}$ is the stress tensor and $f_i$ is the body force acting on a unit volume element $(i,j=x,y)$.  Under an external electromagnetic field, the force is $\bm{f} = \rho e (\bm{E} + \bm{v} \times \bm{B})$.  
The stress tensor complies with the symmetry of the fluid and the absence of energy dissipation.  The latter requires $\Sigma_{ij} (\partial_j v_i) = 0$. \cite{Landau-Lifshitz}  
Then, for a Galilean-invariant fluid without time-reversal-symmetry, the stress tensor has the form 
$\Sigma_{ij} = -P \delta_{ij} 
+ (\eta_\mathrm{H} / 2) [ \epsilon_{ik}(\partial_j v_k + \partial_k v_j) + \epsilon_{jk}(\partial_i v_k + \partial_k v_i) ]$, \cite{Avron1,Read-Rezayi}
where $\epsilon_{ij}$ is the Levi--Civita symbol in two dimensions.  
$P$ and $\eta_\mathrm{H}$ are the pressure and the Hall viscosity, respectively.  
They may be functions of isotropic scalar quantities, which allows gradient expansions.

We solve Equation~\ref{eq:EOM} with $l \gg l_\mathrm{B}$ and $\omega \ll \omega_\mathrm{c}$ under the uniform external magnetic field.  Here, $l$ is the characteristic length of the system, e.g., the radius of the flow in Figure~1B.  
We may regard the conditions as coarse graining, necessary for the hydrodynamic description.  Equation~\ref{eq:EOM} results in the equation of motion (see Methods)
\begin{equation}
m^\ast (\partial_t + \bm{v} \cdot \nabla) \bm{v} = e (\bm{E} + \bm{v} \times \bm{B}) 
- \hbar \chi \Delta \bm{v} \times \hat{z}, 
\label{eq:hydrodynamic}
\end{equation}
where $\chi = K/(\rho\hbar\omega_\mathrm{c}) - \eta_\mathrm{H}/(\rho\hbar)$ is the dimensionless constant with the bulk modulus $K = \rho dP/d\rho$: $\chi = 3N/4$ for an integer quantum Hall state with $\nu = N$ \cite{Hoyos} and 
$\chi = (2k-3)/4$ for a Laughlin state with $\nu = 1/(2k+1)$. \cite{Tokatly3,Read-Rezayi} 

We study the electron flow in Figure~1B.  
We focus on the flow in the bulk by imposing the ``slip'' boundary condition on the walls: the tangential component of the current is unrestricted, and the normal component vanishes since the fluid has no dissipation.  
In a steady state $(\dot{\bm{v}} = \bm{0})$ with the rotational symmetry preserved, the radial velocity $v_r$ should vanish with $\sigma_{xx} = 0$ and only the azimuthal velocity $v_\theta$ can be finite.  
Then, the radial component of Equation~\ref{eq:hydrodynamic} becomes 
\begin{align}
- m^\ast \frac{v_\theta^2}{r} 
= F_r 
- \hbar \chi \frac{\partial}{\partial r} \left( \frac{1}{r} \frac{\partial}{\partial r} (r v_\theta) \right), 
\label{eq:Navier-Stokes}
\end{align}
where $F_r = eE_r + eBv_\theta$ is the force exerted on an electron in the radial direction.  
On the left-hand side, $m^\ast \rho v_\theta^2 / r$ describes the centrifugal force, originating from the convection term of the Navier--Stokes equation.  Importantly, it depends quadratically on $v_\theta$, which makes the equation nonlinear.

We constrain the radial electric field $E_r = a/r$ ($a$: constant), which is realized inside a cylindrical capacitor.  
The vanishing longitudinal conductivity $\sigma_{xx} = 0$ restricts the azimuthal current $j_\theta$ orthogonal to the radial electric field $E_r$ at every point.  
We can solve Equation~\ref{eq:Navier-Stokes} with the series expansion about $r^{-1}$.  
Here, the radius $r$ sets the length scale and we require $r \gg l_\mathrm{B}$.  
Putting $v_\theta = -\sum_{n=1}^{\infty} c_n / r^n$, we find the recurrence relation for the coefficient $c_n$ (Supplemental Section 4).  
In calculating the Hall current $j_\theta = e \rho v_\theta$, we notice the nonuniform density $\rho$ from finite vorticity $\omega = \sum_{n=2}^\infty (n-1) c_n / r^{n+1}$.  
As a result, we obtain the nonlinear current response 
\begin{align}
j_\theta 
&= -\frac{\nu e^2}{h} E_r 
\left( 1 + \frac{E_r}{\mathcal{E}_r} + \frac{4E_r^2}{\mathcal{E}_r^2} + \frac{15E_r^3}{\mathcal{E}_r^3} \right)
+ O(r^{-9}), 
\label{eq:result}
\end{align}
where we define the characteristic electric field strength 
\begin{equation}
\label{eq:characteristic}
\mathcal{E}_r = \frac{e B^2 r}{m^\ast}. 
\end{equation}
The nonlinear components scale with the ratio $E_r / \mathcal{E}_r$ (see also Discussion).  
The effects of the pressure gradient and the Hall viscosity, represented by $\chi$, give corrections on order of $\chi (l_\mathrm{B}/r)^2$ (Supplemental Section 4).   

The result of Equation~\ref{eq:result} is consistent with the quantized linear Hall conductivity $\sigma_\mathrm{H} = \nu e^2/h$ at linear response, and any nonlinearity vanishes for a straight flow with $r \to \infty$.  
The topological protection of $\sigma_{xy}$ remains intact after an adiabatic deformation of the current path without Landau level mixing.  In contrast, the nonlinear components are not topologically protected, and as such, they depend on $r$.  
The nonlinear response originates from the centrifugal force exerted on a curved flow and the density gradient caused by vorticity.  
We may dub Equation~\ref{eq:result} a \textit{nonlinear curvelet} of the curvature $\kappa = r^{-1}$ as it is a fundamental solution under $E_r \propto r^{-1}$, bearing a nonlinear current-voltage relation. 
We note that the hydrodynamic equation (Equation~\ref{eq:hydrodynamic}) is compatible with Laughlin's argument as it concludes no nonlinear response with the configuration in Figure~1A.

\begin{figure}
\centering
\includegraphics[width=\hsize]{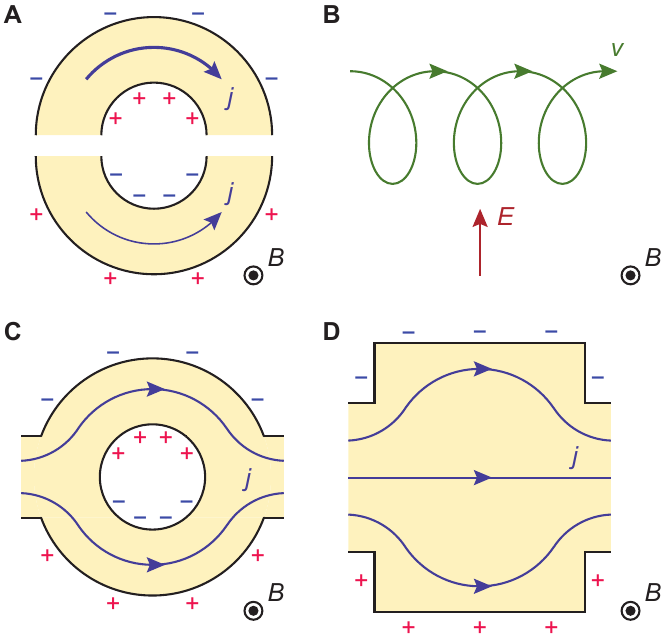}
\caption{%
\textbf{Geometry effects for quantum Hall fluids}\\
(A) Quantum Hall channels with opposite curvatures.  $+$ and $-$ indicate the charge distribution for the flow to bend. 
\newline
(B) Classical trajectory of a cyclotron motion with a drift.  
\newline
(C) Quantum Hall fluid flow in a Mach--Zehnder interferometer. 
\newline
(D) Quantum Hall channel with the rectangular notches.  
}
\label{fig:flow}
\end{figure}

\subsection*{Geometry effects in a quantum Hall flow} 

In deriving Equation~\ref{eq:result}, we fix the electric field $E_r$ to solve Equation~\ref{eq:Navier-Stokes} and calculate the current response.  In an experiment, on the contrary, we often apply a current and measure the Hall voltage, corresponding to distinct boundary conditions.  
In Figure~2A, we consider curved electron channels and apply currents through cross sections of a Corbino disk.  
A transient state would develop a density gradient in the bulk as well as a charge accumulation along the edges.  Upon reaching a steady state, the system supports a curved current flow with a radial electric field. \cite{Schade,Gentile}  
A line integral of the electric field between the inner and outer edges leads to the Hall voltage $V_\mathrm{H} = -\int_\ell \bm{E} \cdot d\bm{\ell}$.  
We here do not discuss edge transport via the compressible one-dimensional states formed by the confining potential, \cite{edge} which may be relevant in Hall bar devices that support essentially straight current flow. \cite{He}

Suppose that we tune the inlet currents to observe the same Hall voltage in the two oppositely-curved channels in Figure~2A.  Though the electric field distribution would not exactly be proportional to $1/r$, Equation~\ref{eq:result} infers a stronger current in the top image of Figure~2A with $E_r > 0$ for $e > 0$.  
The differences appear at even-order terms in $E_r$, which are also sensitive to the sign of the carrier charge through $\mathcal{E}_r$.  
The classical trajectory of a cyclotron motion (Figure~2B) helps an intuitive understanding of the result: the curvature of the flow that complies with the cyclotron motion results in a stronger current.

Aiming at more realistic situations, we discuss a quantum Hall fluid in a configuration similar to a Mach--Zehnder interferometer.  
In Figure~2C, we apply an electric current externally to the quantum Hall system.  
The current from the left splits into two paths.  After traversing either path, it merges and flows out of the Corbino disk.  
The charge density distribution that invokes an electric field reflects the sample geometry and the steady flow.  We assume here that the current injected to the system $I$ is given and that the current flow in the system determines the Hall voltage $V_\mathrm{H}$.  In contrast, an attachment of a long straight channel before injection alters the boundary condition, where the linear $I$-$V_\mathrm{H}$ relation developed in the straight channel would succeed.

As the flow splits, the electric field distribution resembles that of two series capacitors, which have opposite curvatures.  
We suppose no net charge in the inner circle.  
The outlet flow through the channel is the sum of the upper and lower flows with the current density described by Equation~\ref{eq:result}.  Considering the directions of the electric field and the current, we find that the even-order terms are destructive in the discharge.  Therefore, the lowest-order nonlinear response appears at $E_r^3$.  The cancellation of the even-order terms is the consequence of inversion symmetry in the absence of the current.

We deform the channel shape into the one with rectangular notches as shown in Figure~2D.  It does not have a hole that disturbs straight flows like Figure~2C.  
As the charge distribution should reflect the channel widening in the notched region, the nonuniform electric field dilates the flow similarly to Figure~2C, inducing the vorticity, particularly near the corners. \cite{Fujimoto}  Without a hole, there must be no singularity at the center; we note that Equation~\ref{eq:Navier-Stokes} is solvable with $E_r \propto r$ (Supplemental Section 4).  
The total current $I$ passing through any cross section is conserved because of the continuity.  Thus, $I = \int_\ell \bm{j} \cdot \bm{n} \, d\ell$ is constant, where $\bm{j} = e\rho \bm{v}$ is the current density and $\bm{n}$ is the vector normal to the cross section $\ell$.  
On the other hand, the Hall voltage $V_\mathrm{H}$ would reveal a nonlinear relation to the current $I$ reflecting the nonuniform electric field and the flow dilation, which we can infer from the previous discussion for Figure~2C.

\section*{DISCUSSION}

We have shown that a certain nonuniform electric field induces a nonlinear response in the quantum Hall state.  Our theoretical description applies to both integer and fractional quantum Hall effects with bulk energy gaps, where the longitudinal conductivity vanishes.  
We may extend the analysis to Chern insulators, which share some essential properties with the quantum Hall effect.  
We assume the axially symmetric configuration to obtain the analytic solution.  
The Hall current depends on the sample geometry through boundary conditions, which confine the charge distribution and thereby determine the electric field.
Strictly speaking, we should solve the equation of motion (Equation~\ref{eq:EOM}) and Maxwell's equations simultaneously to determine the electric charge distribution and the current response.  

Hydrodynamic electron flows have been studied in high-mobility systems with electron-electron interaction leading to a dissipative viscosity, \cite{Gurzhi1,Molenkamp,Levitov,hydrodynamics3} which is absent in a quantum Hall state.  
The shape effect is discussed also for diffusive electronic transport \cite{Schade} along with experiments, \cite{Makushko} where electric current in a curved path nonlinearly creates the transverse potential.  In a quantum Hall state, we expect the same effect, and moreover, we can reverse the roles of the potential and the current.  As their relative directions are strictly tied with the incompressible bulk, even a nonuniform electric field generates nonlinear current response.  
Observations of quantum Hall states in a Corbino geometry are reported, \cite{Dolgopolov,Jeanneret,Wiegers,Dean} whereas there has been no report yet of a nonlinear Hall response to our knowledge.  
Considering ballistic transport instead, one may find a similarity to the Magnus Hall effect in a time-reversal-invariant noncentrosymmetric material. \cite{Papaj} It, however, requires a built-in electric field in addition to the voltage difference for driving current.  

While we have focused on curved flows, nonuniformity has a relation to nonlinear responses, \cite{Altshuler,Landauer} implying a connection to finite-size effects. \cite{Jeckelmann} 
In quantum Hall systems, Galilean invariance allows finite wavelength corrections to the quantized Hall conductivity $\sigma_\mathrm{H}$ from the Hall viscosity, \cite{Hoyos} where a spatially varying electric field causes a shear flow.  A density modulation induced by finite vorticity could result in nonlinear response at finite wavevectors, but oscillatory contributions do not change the Hall voltage $V_\mathrm{H}$ or the net current $I$ (Supplemental Section 3).  

There is a tiny radiative correction to the Hall conductivity of the integer quantum Hall effect from quantum electrodynamics \cite{Penin} while that from gravitation is absent. \cite{Gravitation}  
In the present analysis, we have neglected radiation of electromagnetic waves by the acceleration of electrons as we suppose low electron velocities.

The nonlinear response that we have discussed does not alter the topologically-protected linear Hall conductivity $\sigma_\mathrm{H}$ but gives rise to a nonlinear $I$-$V_\mathrm{H}$ relation.  A measurement with a \textit{finite} bias (current or voltage) thus observes $I / V_\mathrm{H} \neq \sigma_\mathrm{H}$.  
We may use this property as an experimental probe for sample characterizations.  
We note that optical excitations, where transitions across Landau levels induce nonlinear response, \cite{Avetissian,opticalFQHE} are absent here. 
Throughout the discussion, we have considered the situation without a quasiparticle excitation.  A vortex-like excitation might also induce nonlinear response, \cite{Burgress} while it destroys the quantum Hall state.

To estimate the magnitude of nonlinear response, we evaluate the characteristic electric field strength $\mathcal{E}_r$ in Equation~\ref{eq:characteristic}.  Since $\mathcal{E}_r \propto r$, a flow with higher curvature is preferable in observing the nonlinear effect.  
With $m^\ast$ and $e$ being the bare electron mass and charge, we have $\mathcal{E}_r = 1.76 \times 10^7\,\text{V/m}$ at $B = 10\,\text{T}$ and $r = 1\,\text{{\textmu}m}$, when the magnetic length is $l_\mathrm{B} = 8.1\,\text{nm}$.    
For resistance metrology based on the quantum Hall effect, much wider devices on order of 100\,{\textmu}m to 1\,mm are typically used, \cite{Poirier} making the nonlinearity virtually invisible.  
It is perspicacious to compare the characteristic electric field $\mathcal{E}_r$ with the electric field at the breakdown of the quantum Hall effect. \cite{Eaves,Tsemekhman2,Nachtwei} 
Though its exact value varies depending on mechanisms and details, it is of the order of $E_\mathrm{c} \sim \hbar \omega_\mathrm{c} / (e l_\mathrm{B})$, which leads to $\mathcal{E}_r \sim (r/l_\mathrm{B}) E_\mathrm{c}$, indicating that $\mathcal{E}_r$ is larger than $E_\mathrm{c}$.  
However, this does not mean that the nonlinear response is unobservable, but it does imply that it is detectable near the breakdown, in particular, given the precision of quantum Hall measurements.

In Equation~\ref{eq:result}, we have made an expansion in terms of the inverse radius $r^{-1}$.  It corresponds to the curvature of a streamline $\kappa = R^{-1}$, where $R$ is the radius of the osculating circle, and finite $\kappa$ induces nonlinear response.  
Since the odd-order responses in $E_r$ does not depend on the sign of the curvature $\kappa$, we infer that a meandering current path with alternating $\kappa$ caused by e.g., sample shape and strong inhomogeneity results in nonlinear response.

We can derive Bernoulli's law from the hydrodynamic equation (Equation~\ref{eq:hydrodynamic}) for a steady flow \cite{Avron2} (Supplemental Section 2).  As a result, the velocity $v$, the vorticity $\omega$, and the electric potential $\Phi$ are constant along a streamline.  
The vorticity is affected by the sample geometry and the electric field whereas the Hall viscosity or a pressure gradient is not a source of vorticity. 
The constant electric potential along a streamline is a consequence of the vanishing longitudinal conductivity, which imposes $\bm{v} \perp \bm{E} \parallel \nabla \Phi$.

\section*{METHODS}

\subsection*{Hydrodynamic equations of motion}

The flow of a quantum Hall fluid is governed by the constraint on the density (Equation~\ref{eq:density_constraint}), the momentum equation (Equation~\ref{eq:EOM}), and the condition of an incompressible fluid $D\rho / Dt = 0$ (or equivalently $\nabla \cdot \bm{v} = 0$).  
Strictly speaking, we should also take account of Maxwell's equations self-consistently.

Here, we look into the stress tensor $\Sigma_{ij}$ in detail.  
It takes the form 
\begin{align}
\Sigma_{ij} = -P \delta_{ij} + \Sigma'_{ij}, 
\label{eq:stress_tensor}
\end{align}
where $P$ is the pressure and $\Sigma'_{ij}$ is the viscosity stress tensor.  As a rank-two tensor, one may in general write the viscosity stress tensor as $\Sigma'_{ij} = \eta_{ijkl} \textit{\textsf{E}}_{kl}$, where $\eta_{ijkl}$ is the viscosity tensor and $\textit{\textsf{E}}_{ij} = ( \partial_i v_j + \partial_j v_i ) /2$ is the strain rate tensor. 

The form of the viscosity tensor must respect the symmetry of the system.  We here impose Galilean invariance on the fluid while time-reversal symmetry is broken by the external magnetic field.  
In an isotropic two-dimensional system, the viscosity tensor $\eta_{ijkl}$ has only one independent component for the antisymmetric part $\eta^{\mathrm{(A)}}_{ijkl} = -\eta^{\mathrm{(A)}}_{klij}$, which invokes no dissipation. \cite{Avron2}  We refer to it as the Hall viscosity and denote it as $\eta_\mathrm{H}$.  (Terminologies vary depending on literature.  We note that $\eta_\mathrm{H}$ is the dynamic Hall viscosity, which in SI has the unit $[\eta_\mathrm{H}] = \mathrm{kg\,s^{-1}}$, while the kinematic Hall viscosity $\nu_\mathrm{H}$ has the unit $[\nu_\mathrm{H}] = \mathrm{m^2\,s^{-1}}$.)
With only the Hall viscosity present, the viscosity tensor in the Cartesian coordinate is \cite{Avron1,Read-Rezayi}
\begin{equation}
\eta_{ijkl} = \eta_\mathrm{H} (\delta_{il} \epsilon_{jk} + \delta_{jk} \epsilon_{il}).  
\end{equation}
We note that $\eta_\mathrm{H}$ and $eB$ have the same sign. 

With the viscosity tensor given, the stress tensor becomes 
\begin{align}
\Sigma_{ij} &= -P \delta_{ij} 
+ \frac{\eta_\mathrm{H}}{2}[ \epsilon_{ik}(\partial_j v_k + \partial_k v_j) + \epsilon_{jk}(\partial_i v_k + \partial_k v_i) ] \nonumber\\
&= -P \delta_{ij} 
+ \frac{\eta_\mathrm{H}}{2} (\partial_j v_i^\ast + \partial_i^\ast v_j + \partial_i v_j^\ast + \partial_j^\ast v_i), 
\label{eq:S_stress-tensor}
\end{align}
where we define the star notation for a dual vector $a_i^\ast = \epsilon_{ij} a_j$ ($\bm{a}^\ast = \bm{a} \times \hat{z}$).  
The pressure $P$ and the Hall viscosity $\eta_\mathrm{H}$ are scalar quantities, which may further contain gradient expansions.  Nonetheless, the effect of the stress tensor is perturbatively small in considering the nonlinear response as we will show later, and thus we regard $\eta_\mathrm{H}$ as a constant.

\subsection*{Slowly varying conditions}

We introduce some approximations to proceed with the discussion, which separate the microscopic and macroscopic scales of the system.  
We write the length scale of the velocity or density gradient as $l$ and the frequency scale of a steady flow is set by the vorticity $\omega$.  
Now we recall that the characteristic length and frequency scales of a quantum Hall state are the magnetic length $l_\mathrm{B} = \sqrt{\hbar / (eB)}$ and the cyclotron frequency $\omega_\mathrm{c} = eB / m^\ast$, respectively. 
We thus assume 
\begin{equation}
\label{eq:S_slowly-varying}
l \gg l_\mathrm{B}, \quad 
\omega \ll \omega_\mathrm{c}.  
\end{equation}
Those conditions are usually satisfied; otherwise, the quantum Hall state would be destroyed or the hydrodynamic description is invalid.  
In other words, we distinguish the intrinsic \textit{microscopic} scales of an electronic system under a magnetic field, i.e., $l_\mathrm{B}$ and $\omega_\mathrm{c}$, from the the \textit{macroscopic} scales of a flow $l$ and $\omega$.

\subsection*{Derivation of the equation of motion}

Now we derive the equation of motion for $\bm{v}$ (Equation~\ref{eq:hydrodynamic}) from the momentum equation (Equation~\ref{eq:EOM}).  
We aim at obtaining the equation of motion without explicit dependence on the density $\rho$.  
We assume the slowly varying conditions (Equation~\ref{eq:S_slowly-varying}) and a \textit{uniform} external magnetic field.  Then, it is instructive to write Equation~\ref{eq:density_constraint} as 
\begin{equation}
\label{eq:S_density-3}
\rho = \bar{\rho} \left( 1 + \frac{\omega}{\omega_\mathrm{c}} \right) \quad 
\mathrm{with}\quad \bar{\rho} = \frac{\nu e B}{h}.  
\end{equation}

We evaluate the derivative of the stress tensor $\partial_j \Sigma_{ij}$.  We suppose that the pressure $P$ and the Hall viscosity $\eta_\mathrm{H}$ depend on the density $\rho$.  
First we look at the gradient of the pressure, using the relation $\nabla P = (\nabla \rho) (\partial_\rho P)$.  
With the density gradient $\nabla \rho = \bar{\rho} (\nabla \omega) / \omega_\mathrm{c}= (\bar{\rho} / \omega_\mathrm{c}) \Delta \bm{v}^\ast$ and the definition of the bulk modulus $K = \rho \partial P(\rho) / \partial\rho$, we obtain 
\begin{equation}
\label{eq:S_pressure-gradient}
\nabla P 
= \frac{K}{\omega_\mathrm{c}} \Delta \bm{v}^\ast, 
\end{equation}
where we consider weak density modulation $\rho \approx \bar{\rho}$ for $\omega \ll \omega_\mathrm{c}$.  

Next we calculate the viscosity term
\begin{align}
&\quad \partial_j \Sigma'_{ij} \nonumber\\
&= (\delta_{il} \epsilon_{jk} + \delta_{jk} \epsilon_{il}) \partial_j (\eta_\mathrm{H}  \textit{\textsf{E}}_{kl}) \nonumber\\
&= (\delta_{il} \epsilon_{jk} + \delta_{jk} \epsilon_{il}) \eta_\mathrm{H} (\partial_j \textit{\textsf{E}}_{kl})
+ (\delta_{il} \epsilon_{jk} + \delta_{jk} \epsilon_{il}) (\partial_j \eta_\mathrm{H}) \textit{\textsf{E}}_{kl}.  
\end{align}
It is straightforward to show that the first term reduces to $\eta_\mathrm{H} \Delta v^\ast_i$.  
We argue that the second term is negligible under the present assumptions.  
When the Hall viscosity shows polynomial dependence on $\rho$ as we will see in Suppelementary Information, we find $\partial_j \eta_\mathrm{H} = (\partial_j \rho) (\partial_\rho \eta_\mathrm{H}) \approx (\partial_j \rho) (\eta_\mathrm{H} / \rho)$.  Then, we can estimate the magnitude of the second term as $\eta_\mathrm{H} \rho^{-1} |\nabla \rho| |\textit{\textsf{E}}|$, so that the ratio to the first term is $\eta_\mathrm{H} \rho^{-1} |\nabla \rho| |\textit{\textsf{E}}| / (\eta_\mathrm{H} \Delta \bm{v}^\ast) \sim |\textit{\textsf{E}}| / \omega_\mathrm{c}$.  
For $\omega \ll \omega_\mathrm{c}$, this ratio should be negligibly small, and hence we omit the Hall viscosity gradient term.  

At last, considering the body force $\bm{f} = \rho e (\bm{E} + \bm{v} \times \bm{B})$ in the presence of the external electric and magnetic fields, we arrive at Equation~\ref{eq:hydrodynamic} and identify the dimensionless quantity $\chi = K/(\rho\hbar\omega_\mathrm{c}) - \eta_\mathrm{H}/(\rho\hbar)$.  
The equation of motion now consists only of the velocity $\bm{v}$.  After we determine the velocity $\bm{v}$, we can deduce the density $\rho$ from Equation~\ref{eq:density_constraint}.

\subsection*{Acknowledgments}
We would like to thank P. He for experimental data and N. Nagaosa for discussions.  We are also grateful to M. Kawamura for informative conversations. 
This work was supported by JSPS KAKENHI Grant Number JP24H00197.

\clearpage
\onecolumngrid

\begin{center}
\textbf{\large Supplemental Methods}
\end{center}

\setcounter{section}{0}
\setcounter{equation}{0}
\setcounter{figure}{0}
\renewcommand{\thesection}{\arabic{section}}
\renewcommand{\theequation}{S\arabic{equation}}
\renewcommand{\thefigure}{S\arabic{figure}}

\renewcommand{\figurename}{\textbf{Figure}}
\renewcommand{\thefigure}{\textbf{S\arabic{figure}}}

\renewcommand\refname{Supplemental references}

\section{Lorentz invariance of a quantum Hall state}

We consider a uniform two-dimensional electron gas on the $xy$ plane and apply a uniform external magnetic field $\bm{B} = B\hat{z}$ along the $z$ direction.  We prepare two inertial frames: in one frame, electrons are at rest.  To be more precise, the total momentum of electrons vanishes.  We first sit in the rest frame.  
When the electronic system is in a quantum Hall state, the electron density $\rho$ satisfies 
\begin{equation}
\rho = \frac{\nu e B}{h}.  
\end{equation}
As per the assumption, the current density $\bm{j}$ should vanish: $\bm{j} = \bm{0}$.  

Then, we shift our focus to the moving frame, which moves relative to the rest frame at velocity $-\bm{v} = -v_x \hat{x}$.  The electrons now move at velocity $+\bm{v}$.  
In the moving frame, the length parallel to the velocity $\bm{v}$, i.e., the $x$ direction, shrinks because of the Lorentz contraction.  It increases the density $\rho$ to be 
\begin{equation}
\rho' = \frac{\rho}{\sqrt{1-v_x^2/c^2}}, 
\end{equation}
where $c$ is the speed of light and primes in this section indicate quantities in the moving frame.  
The Lorentz transformation also modifies the electromagnetic fields.  Though the electromagnetic fields are purely magnetic in the rest frame, the electric and magnetic fields in the moving frame $\bm{E}'$ and $\bm{B}'$, respectively, are 
\begin{equation}
\bm{E}' = - \frac{\bm{v} \times \bm{B}}{\sqrt{1 - v_x^2/c^2}},  \quad 
\bm{B}' = \frac{\bm{B}}{\sqrt{1 - v_x^2/c^2}}.  
\end{equation}
We recall $\bm{E} = \bm{0}$ in the rest frame.  
With $\bm{v} = v_x \hat{x}$ and $\bm{B} = B\hat{z}$, we find the finite components 
\begin{equation}
E'_y = \frac{v_x B}{\sqrt{1 - v_x^2/c^2}}, \quad 
B'_z = \frac{B}{\sqrt{1 - v_x^2/c^2}}. 
\end{equation}
Considering that the electrons move at velocity $\bm{v}$, we obtain the current density along the $x$ direction
\begin{equation}
j'_x = e \rho' v_x = e \frac{\rho}{B} E'_y = \frac{\nu e^2}{h} E'_y.  
\end{equation}
Therefore, the longitudinal and Hall conductivities, $\sigma_{xx}$ and $\sigma_{xy}$, respectively, are  
\begin{equation}
\sigma_{xx} = 0, \quad 
\sigma_{xy} = \frac{\nu e^2}{h}, 
\end{equation}
and importantly, they are Lorentz invariant.  Lorentz invariance requires the longitudinal conductivity $\sigma_{xx}$ to vanish.  

We recall that a Lorentz transformation preserves 
\begin{equation}
c^2 B^2 - E^2 \quad \text{and} \quad \bm{E} \cdot \bm{B}.  
\end{equation}
Since $E = 0$ and $B \neq 0$ in the rest frame, we always have $c^2 B^2 - E^2 > 0$, meaning that we cannot annihilate the magnetic field.  If the electromagnetic fields would be purely electric, a quantum Hall state could not be realized and the electric field would incessantly accelerate the electrons.  The latter condition results in $\bm{E} \cdot \bm{B} = 0$ in the present case; i.e., the electric and magnetic fields are always orthogonal.

\section{Properties of quantum Hall fluids}

In this section, we discuss the properties of quantum Hall fluids before proceeding to examine flows under an external electric field.

\subsection{Incompressibility}

The conservation of the electron number leads to the continuity equation
\begin{equation}
\label{eq:S_continuity}
\partial_t \rho + \nabla \cdot (\rho \bm{v}) = 0, 
\end{equation}
where $\rho$ is the electron density and $\bm{v}$ is the electron velocity.  
Since a quantum Hall state is incompressible with an energy gap, we suppose that the material derivative of the density vanishes: 
\begin{equation}
\label{eq:S_incompressibility}
\frac{D\rho}{Dt} = \partial_t \rho + \bm{v} \cdot \nabla \rho = 0.  
\end{equation}
We can thus rewrite the continuity equation (Equation~\ref{eq:S_continuity}) as 
\begin{align}
\partial_t \rho + \nabla \cdot (\rho\bm{v})
&= \frac{D\rho}{Dt} + \rho (\nabla \cdot \bm{v}) \nonumber\\
&= \rho (\nabla \cdot \bm{v}) \nonumber\\
&= 0.  
\end{align}
Therefore, the velocity field is divergence-free 
\begin{equation}
\label{eq:S_divergence-free}
\nabla \cdot \bm{v} = 0, 
\end{equation}
when the density $\rho$ is finite.  
We emphasize, however, that the density may not be uniform. 
The condition for an isochoric state $\rho = \text{constant}$ is stricter than the that for incompressibility, which is not always the case for a quantum Hall fluid.  We do not assume the isochoric condition in the entire discussion but suppose the divergence-free condition.

\subsection{Stream function}

With the divergence-free condition $\nabla \cdot \bm{v} = 0$, we can immediately define the stream function $\Psi$ that satisfies 
\begin{equation}
v_x = \partial_y \Psi, \quad v_y = -\partial_x \Psi.  
\end{equation}
Equivalently, we can write it in a concise form 
\begin{equation}
\bm{v} = \nabla^\ast \Psi, 
\end{equation}
where we define the star notation for a dual vector
\begin{equation}
a_i^\ast = \epsilon_{ij} a_j, \quad 
\bm{a}^\ast = \bm{a} \times \hat{z}.  
\end{equation}
Here, $\epsilon_{ij}$ is the Levi--Civita symbol in two dimensions, and $i$, $j$ are used for spatial components $(1, 2)$ or $(x, y)$ hereafter.  
This operation corresponds to the rotation about the $z$ axis by $-\pi/2$.  
It is easy to confirm $(\bm{a}^\ast)^\ast = -\bm{a}$.  
(Obviously, we should not confuse the effective mass $m^\ast$ with the star notation.)

The vorticity in terms of the stream function is 
\begin{equation}
\omega = \nabla \times \bm{v} = - \Delta \Psi.  
\end{equation}
Here, we identify $\nabla \times \bm{v}$ as $\hat{z} \cdot (\nabla \times \bm{v})$, since the curl of the two-dimensional vector field has a finite component only along the $z$ direction.  We hereafter use the curl operator $\nabla \times$ acting on a two-dimensional vector field to indicate its $z$ component as we analyze electrons confined on the $xy$ plane.  

Regarding the vorticity, we find the relation 
\begin{equation}
\label{eq:S_velocity-vorticity}
\Delta \bm{v}^\ast = - \Delta (\nabla \Psi) = \nabla (-\Delta \Psi) = \nabla \omega. 
\end{equation}
A quantum Hall fluid may have finite vorticity, and hence it is not a potential flow.  We cannot introduce the velocity potential $\varphi$ that satisfies $\bm{v} = \nabla\varphi$.

\subsection{Dynamics of the vorticity}

From the curl of Equation~5, we obtain the equation of motion for the vorticity 
\begin{equation}
m^\ast \partial_t \omega + m^\ast (\bm{v}\cdot\nabla) \omega = - e [ \partial_t B + (\bm{v}\cdot\nabla) B ]. 
\end{equation}
We use Maxwell's equations in the derivation.  We note that the pressure and Hall viscosity terms vanish as $\nabla \times \Delta \bm{v}^\ast = \nabla \times \nabla \omega = 0$. 
With the material derivative, we can write the equation in a concise form 
\begin{equation}
\label{eq:S_vorticity_EOM}
\frac{D}{Dt} (\omega_\mathrm{c} + \omega) = 0.  
\end{equation}

The result implies that the sum of the cyclotron frequency and the vorticity is conserved as a volume element moves.  We note that the cyclotron frequency depends on the local magnetic field.  When the magnetic field is uniform and constant, the vorticity is constant as we move with the fluid.

\subsection{Bernoulli's law}

In this subsection, we consider a \textit{steady state} with $\partial_t \bm{v} = \bm{0}$ and transform the equation of motion (Equation~5).  
We utilize the identity 
\begin{equation}
\label{eq:S_vector}
f_j \partial_j f_i 
= \frac{1}{2} \partial_i (f_j f_j) - \epsilon^{ij} f_j \epsilon^{kl} \partial_k f_l , 
\end{equation}
which is analogous to the equality in the three-dimensional vector calculus $(\bm{f} \cdot \nabla) \bm{f} = \frac{1}{2} \nabla \bm{f}^2 - \bm{f} \times (\nabla \times \bm{f})$.  
Then, Equation~5 becomes 
\begin{equation}
\label{eq:S_EOM_steady}
m^\ast \left[ \frac{1}{2} \nabla v^2 - \bm{v} \times (\nabla \times \bm{v}) \right] 
= e (\bm{E} + \bm{v} \times \bm{B}) - \hbar \chi \nabla \omega.  
\end{equation}
We denote the electromagnetic potentials as $(\Phi, \bm{A})$, and the electric and magnetic fields $\bm{E}$ and $\bm{B}$, respectively, are 
\begin{gather}
\bm{E} = -\nabla \Phi - \partial_t \bm{A}, \\
\bm{B} = \nabla \times \bm{A}.  
\end{gather}
In a steady state, we may write the electric field as $\bm{E} = -\nabla \Phi$.  
When there is a position-dependent potential $\varphi(\bm{r})$, we have an additional force term $-\nabla \varphi(\bm{r})$ on the right-hand side of Equation~\ref{eq:S_EOM_steady}, which may be included in $e\Phi(\bm{r})$.  

By collecting the gradient terms, we obtain 
\begin{align}
\nabla \left( \frac{1}{2} mv^2 + \chi \hbar \omega + e\Phi \right) 
&= \bm{v} \times [ e\bm{B} + m (\nabla \times \bm{v}) ] \nonumber\\
&= \bm{v}^\ast (eB + m\omega), 
\label{eq:S_Bernoulli-prep}
\end{align}
where we suppose that $\chi$ is constant.  
Since the vector on the right-hand side of the equation is always orthogonal to the flow velocity $\bm{v}$, we find Bernoulli's equation
\begin{equation}
\label{eq:S_Bernoulli1}
\frac{1}{2} mv^2 + \chi \hbar \omega  + e\Phi = \text{constant along a streamline.}
\end{equation}

We can rewrite Equation~\ref{eq:S_Bernoulli-prep} as 
\begin{align}
\nabla \left( \frac{1}{2} mv^2 + \chi \hbar \omega \right) 
&= e\bm{E} + \bm{v}^\ast (eB + m\omega). 
\end{align}
We recall that the longitudinal conductivity vanishes in a quantum Hall state ($\sigma_{xx} = 0$), so that the electric current $\bm{j} = e \rho \bm{v}$ and hence the velocity $\bm{v}$ are orthogonal to the electric field $\bm{E}$.  Therefore, we observe that the right-hand side is still orthogonal to the velocity $\bm{v}$, and obtain another relation
\begin{equation}
\label{eq:S_Bernoulli}
\frac{1}{2} mv^2 + \chi \hbar \omega = \text{constant along a streamline.}
\end{equation}
We now find from Equations~\ref{eq:S_Bernoulli1} and \ref{eq:S_Bernoulli}
\begin{equation}
\Phi = \text{constant along a streamline.}
\end{equation}
This is actually a direct consequence of the vanishing longitudinal conductivity $\sigma_{xx} = 0$.  The velocity $\bm{v}$ is always orthogonal to the electric field $\bm{E} = -\nabla \Phi$, and hence streamlines coincide with the equipotential lines of $\Phi$.

Under a uniform magnetic field, Equation~\ref{eq:S_vorticity_EOM} indicates that the vorticity $\omega$ does not change along a streamline in a steady state.  
In summary, the velocity $v$, the vorticity $\omega$, and the electric potential $\Phi$ are constant along a streamline in an incompressible quantum Hall fluid.

\subsection{The dimensionless parameter $\bm{\chi}$}
\label{sec:S_parameter}

We calculate the dimensionless parameter $\chi$ for noninteracting systems.  We neglect the spin degeneracy, or equivalently we consider cases with spins fully polarized.  

\subsubsection{Integer quantum Hall state}

For a two-dimensional electronic system with an energy dispersion $\epsilon_{\bm{k}} = k^2/(2m^\ast)$, the energy eigenvalues of the Landau levels under a uniform magnetic field $B$ are 
\begin{equation}
\epsilon_n = \left( n + \frac{1}{2} \right) \hbar \omega_\mathrm{c} \quad (n=0,1,2,\cdots).  
\end{equation}
When the $N$ lowest Landau levels are filled, the energy density is 
\begin{equation}
U = \frac{1}{2\pi l_\mathrm{B}^2} \sum_{n=0}^{N-1} \epsilon_n
= \frac{N^2 \hbar \omega_\mathrm{c}}{4\pi l_\mathrm{B}^2} 
= \frac{\rho^2 h^2}{4\pi m^\ast}.  
\end{equation}
Using the relation for the pressure $P = \rho \partial U/\partial \rho$, the bulk modulus $K$ is 
\begin{equation}
K = \rho^2 \left( \frac{\partial^2 U(\rho)}{\partial\rho^2} \right)_N 
= \frac{\rho^2 h^2}{2\pi m^\ast} 
= N \rho \hbar \omega_\mathrm{c}.  
\end{equation}
An electron fluid filling the $n$-th Landau level has the Hall viscosity $(\hbar/2)(n+1/2)/(2\pi l_\mathrm{B}^2)$ [S1], leading to the Hall viscosity of the fluid with the lowest $N$ Landau levels filled 
\begin{equation}
\eta_\mathrm{H} = \sum_{n=0}^{N-1} \frac{\hbar}{2} \left( n+\frac{1}{2} \right) \frac{1}{2\pi l_\mathrm{B}^2} 
= \frac{1}{4} N \rho \hbar.  
\end{equation}
Therefore, the dimensionless parameter $\chi$ becomes 
\begin{equation}
\chi = N - \frac{1}{4} N 
= \frac{3}{4} N.  
\end{equation}

\subsubsection{Fractional quantum Hall effect}

When we neglect the effect of interactions, the energy density for a fractional Hall state with the filling factor $\nu < 1$ is 
\begin{equation}
U = 
\nu \left( \frac{1}{2} \hbar \omega_\mathrm{c} \right) \frac{1}{2\pi l_\mathrm{B}^2}
= \frac{1}{2} \rho \hbar \omega_\mathrm{c}, 
\end{equation}
which leads to the bulk modulus 
\begin{equation}
K = \rho^2 \left( \frac{\partial^2 U(\rho)}{\partial\rho^2} \right)_\nu 
= \rho \hbar \omega_\mathrm{c}.  
\end{equation}
For the Laughlin states with the filling factor $\nu = 1/(2k+1)$, the Hall viscosity is [S2]
\begin{equation}
\eta_\mathrm{H} = \frac{1}{4} \nu^{-1} \rho \hbar.  
\end{equation}
Then, we obtain the dimensionless parameter 
\begin{equation}
\chi = 1 - \frac{1}{4\nu} = \frac{3-2k}{4}.  
\end{equation}

\section{Quantum Hall fluid in a straight channel}
\label{sec:S_straight}

In the section, we examine a quantum Hall fluid in a straight channel that flows along the $x$ direction.  
We suppose a uniform magnetic field $\bm{B} = B\hat{z}$ and an electric field $\bm{E}(y) = E(y) \hat{y}$, which may vary along the $y$ direction.  
We assume the translational symmetry along the $x$ direction in a steady state.  
Then, the fluid velocity lies along the $x$ direction, i.e., $\bm{v} = v_x(y) \hat{x}$.  
If the electric field is slowly varying in space, a velocity gradient arises to cause a shear flow.  Finite vorticity induces a density modulation in the incompressible fluid, following Equation~3.

We apply Equation~5 to obtain the equation of motion
\begin{equation}
0 = e [E(y) - v_x B] + \hbar \chi \partial_y^2 v_x(y).  
\end{equation}
As we here consider a simple shear flow, the divergence-free condition $\nabla \cdot \bm{v} = 0$ is automatically satisfied.  
We suppose that the electric field $E(y)$ varies along the $y$ direction at wavenumber $q$.  
With the Fourier transformation, we can easily solve the equation and obtain the velocity 
\begin{equation}
v_x(q) = \frac{E(q)}{B} \left( 1 + \frac{\chi \hbar}{e B} q^2 \right)^{-1}
= \frac{E(q)}{B} \left[ 1 + \chi (ql_\mathrm{B})^2 \right]^{-1}.  
\label{eq:S_velocity-q}
\end{equation}
We note that the velocity $v_x$ has the same wavenumber $q$ as the electric field since the equation of motion is linear both in $v_x$ and $E$.  Also, $v_x$ is linearly dependent on $E$.  
The spatial dependence of the velocity leads to finite vorticity, which in turn results in the density modulation $\delta\rho$ according to Equation~3: 
\begin{equation}
\delta\rho(q) = -iq \frac{\nu m^\ast}{h} v_x(q).  
\label{eq:S_density-q}
\end{equation}
The density modulation is out of phase with the velocity, and the vorticity is larger with a larger velocity gradient or a shear rate.  

Now we suppose that Fourier components of the electric field is finite at wavevector $q$, i.e., $E(y) \propto \operatorname{Re}e^{iqy}$.  
We calculate the electric current along the $x$ direction $j_x = e (\bar{\rho} + \delta\rho) v_x$, where $\bar{\rho} = \nu e B / h$ is the uniform background density; see Equation~13.  
Then, its wavenumber-$q$ component is 
\begin{equation}
j_x (q) = e \bar{\rho} v_x(q).  
\end{equation}
We note that there is no contribution from $\delta\rho$ since $v_x(q=0) = 0$.  
Thus, the linear Hall conductivity of a quantum Hall fluid at filling factor $\nu$ becomes 
\begin{align}
\sigma_{xy}(q) 
&= \frac{e \bar{\rho} v_x(q)}{E(q)} 
= \frac{\nu e^2}{h} \left[ 1 + \chi (ql_\mathrm{B})^2 \right]^{-1} \nonumber\\
&= \frac{\nu e^2}{h} \left[ 1 - \chi (ql_\mathrm{B})^2 + O((ql_\mathrm{B})^4) \right].  
\end{align}
The result reproduces the result by Hoyos and Son to order $(ql_\mathrm{B})^2$ [S3].  

In addition to the linear response, we can think of a nonlinear response since the density modulation Equation~\ref{eq:S_density-q} depends on the velocity and hence the electric field.  Here, we calculate the electric current at wavevectors $2q$ and $0$: 
\begin{gather}
j_x(2q) = e \delta\rho(q) v_x(q) = -iq \frac{\nu m^\ast}{h} v_x^2(q), \\
j_x(0) = e [\delta\rho(q) v_x(-q) + \delta\rho(-q) v_x(q)] = 0.  
\end{gather}
The uniform component does not contain a finite nonlinear response and it vanishes identically.  
We see that the $2q$ component is proportional to the wavenumber $q$. 
Therefore, the finite-$q$ corrections and the nonlinear terms with respect to the electric field vanish in the limit $q \to 0$, which is compatible with Galilean invariance.

\section{Quantum Hall fluid in axially symmetric systems}

We investigate axially symmetric systems in quantum Hall states and solve the hydrodynamic equation (Equation~5). 
We suppose a radial electric field $\bm{E}(r) = E_r(r) \hat{r}$ and an azimuthal velocity $\bm{v}(r) = v_\theta(r) \hat{\theta}$ because of $\sigma_{xx} = 0$ and $\sigma_{xy} \neq 0$.  
We do not consider a spontaneous symmetry breaking of the continuous rotational symmetry and assume that the symmetry is conserved.  We can easily confirm that the radial flow with velocity $v_\theta(r)$ satisfies the divergence-free condition $\nabla \cdot \bm{v} = 0$. 

It is worth recalling that the hydrodynamic equation for a usual viscous fluid is solvable for vortex flows, resulting in a velocity field proportional to $r$ or $r^{-1}$.  
Inspired by this observation, we consider the electric fields $E_r = a / r$ and $E_r = a' r$, where $a$ and $a'$ are constants.  
We constrain our discussion on steady states in this section.

The functions $r$ and $r^{-1}$ diverge at $r \to \infty$ and $r \to +0$, respectively, and boundary conditions are indispensable for the solutions to the differential equation.  Since a quantum Hall fluid has no dissipative viscosity, the no-slip boundary condition, which accompanies a momentum transfer near boundaries, is prohibited.  
We instead suppose the ``slip'' boundary condition: the velocity tangential to the boundary is unrestricted while the normal component should vanish.

The radial component of the hydrodynamic equation (Equation~5) in the polar coordinate is 
\begin{equation}
\label{eq:S_hydrodynamic_polar}
- m^\ast \frac{v_\theta^2}{r}= e (E_r + v_\theta B) - \hbar \chi \partial_r \left( \frac{1}{r} \partial_r (r v_\theta) \right).  
\end{equation}
On the other hand, the azimuthal component vanishes identically under the present conditions.  
We note that the dominant contribution in the equation of motion comes from the electromagnetic fields, namely, the first term of the right-hand side.  
It gives rise to the leading-order response, i.e., the quantized Hall response.  The other two terms give corrections, which we will see in the following.  

For a later use, it is convenient to write the vorticity in the polar coordinate
\begin{equation}
\label{eq:S_vorticity_polar}
\omega = \partial_x v_y - \partial_y v_x = \frac{1}{r} \partial_r (r v_\theta) - \frac{1}{r} \partial_\theta v_r.  
\end{equation}

\subsection{Electric field proportional to $\bm{1/r}$}
\label{sec:S_1/r}

We first consider the external electric field 
\begin{equation}
E_r = \frac{a}{r}.  
\end{equation}
We can realize this electric field inside a concentric cylindrical capacitor (Figure~S1A).  
Suppose that the inner and outer cylinders are oppositely charged with $\pm\lambda$ being the charge densities per height and azimuthal angle.  
Then, the electric field along the radial direction is confined inside the capacitor, which realizes $E_r = \lambda / (\varepsilon r)$, where $\varepsilon$ is the permittivity inside the capacitor (Figure~S1B).  We place a quantum Hall system in the Corbino geometry inside the capacitor.  The radial electric field covers in the three dimensional space while the electronic system is confined in the two dimensional plane, so that the electron flow barely disrupts the radial electric field.  

\begin{figure}
\centering
\includegraphics[width=0.55\hsize]{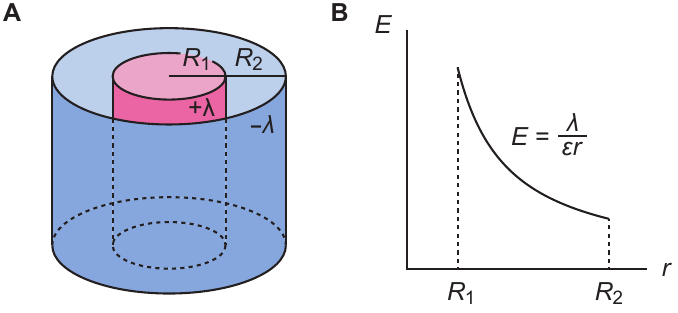}
\caption{%
\textbf{Cylindrical capacitor}
\newline 
(A) Concentric cylindrical capacitor.  $\lambda$ is the charge density per height and azimuthal angle.  
\newline 
(B) Radial electric field inside the capacitor.  $\varepsilon$ is the permittivity inside the capacitor. 
}
\label{fig:field}
\end{figure}

The expansion of the azimuthal velocity 
\begin{equation}
\label{eq:S_velocity_expansion}
v_\theta = - \sum_{n=1}^\infty \frac{c_n}{r^n}
\end{equation}
allows us to solve Equation~\ref{eq:S_hydrodynamic_polar}, which becomes after we substitute $E_r$ and $v_\theta$
\begin{equation}
- m^\ast \sum_{n=3}^\infty \frac{1}{r^n} \sum_{n'=1}^{n-2} c_{n'} c_{n-n'-1} 
= e \frac{a}{r} - eB \sum_{n=1}^\infty \frac{c_n}{r^n} 
+ \hbar \chi \sum_{n=4}^\infty c_{n-2} [(n-2)^2-1] \frac{1}{r^n}.  
\end{equation}
From the equations at orders $r^{-n}$ $(n=1,2,3)$, we obtain 
\begin{gather}
c_1 = \frac{a}{B}, \\
c_2 = 0, \\
c_3 = \frac{m^\ast}{eB} c_1^2 = \frac{a^2}{\omega_\mathrm{c} B^2}.  
\end{gather}
For $n \geq 4$, we utilize the nonlinear recurrence relation
\begin{equation}
c_n = \frac{1}{\omega_\mathrm{c}} \sum_{n'=1}^{n-2} c_{n'} c_{n-n'-1} + \chi l_\mathrm{B}^2 c_{n-2} [(n-2)^2 - 1].  
\end{equation}
The recurrence equation indicates that all $c_n$ for even $n$ vanish by induction.  
Let us prove this statement: 
We can easily check $c_4 = 0$ and then assume $c_{n'} = 0$ for even $n'$ smaller than $n(>4)$.  For even $n$, every term of the summation $c_{n'} c_{n-n'-1}$ consists of the product of even- and odd-order coefficients at orders smaller than $n$, which should vanish by the assumption.  The other term with $c_{n-2}$ should also vanish by assumption.  
Therefore, we can conclude $c_{n} = 0$ for even $n$.  On the other hand, the coefficients $c_n$ are finite for odd $n$.  
By solving the equation to order $r^{-7}$, namely $c_7$, we obtain the velocity 
\begin{align}
v_\theta &= - \sum_{n=1}^\infty \frac{c_n}{r^n} \nonumber\\
&= -\underbrace{\frac{E_r}{B}}_{r^{-1}} 
-\underbrace{\frac{E_r}{B} \frac{E_r}{\mathcal{E}_r}}_{r^{-3}} 
-\underbrace{\frac{E_r}{B} \left( \frac{2E_r^2}{\mathcal{E}_r^2} + 8\chi \frac{E_r}{\mathcal{E}_r} \frac{l_\mathrm{B}^2}{r^2} \right)}_{r^{-5}} 
-\underbrace{\frac{E_r}{B} \left( \frac{5E_r^3}{\mathcal{E}_r^3} + 64\chi \frac{E_r^2}{\mathcal{E}_r^2} \frac{l_\mathrm{B}^2}{r^2} + 192\chi^2 \frac{E_r}{\mathcal{E}_r} \frac{l_\mathrm{B}^4}{r^4} \right)}_{r^{-7}} \nonumber\\
&\quad +\, O(r^{-9}) \nonumber\\
&= -\frac{E_r}{B} \left[ 
1 
+ \frac{E_r}{\mathcal{E}_r} \left( 1 + \frac{8\chi l_\mathrm{B}^2}{r^2} + \frac{192\chi l_\mathrm{B}^4}{r^4} \right) 
+ \frac{2E_r^2}{\mathcal{E}_r^2} \left( 1 + \frac{32\chi l_\mathrm{B}^2}{r^2} \right) 
+ \frac{5E_r^3}{\mathcal{E}_r^3}
\right]
+ O(r^{-9}).  
\end{align}
Here we introduce the characteristic electric field strength 
\begin{equation}
\mathcal{E}_r = \frac{e B^2 r}{m^\ast}.  
\end{equation}

To evaluate the current density, we also need the density distribution.  
For the azimuthal velocity Equation~\ref{eq:S_velocity_expansion}, the vorticity becomes 
\begin{align}
\omega &= \sum_{n=2}^\infty (n-1) \frac{c_n}{r^{n+1}} \nonumber\\
&= \omega_\mathrm{c} \underbrace{\frac{2E_r^2}{\mathcal{E}_r^2}}_{r^{-4}} 
+\, \omega_\mathrm{c} \underbrace{\left( \frac{8E_r^3}{\mathcal{E}_r^3} + 32\chi \frac{E_r^2}{\mathcal{E}_r^2} \frac{l_\mathrm{B}^2}{r^2} \right)}_{r^{-6}} 
+\, \omega_\mathrm{c} \underbrace{\left( \frac{30E_r^4}{\mathcal{E}_r^4} + 384\chi \frac{E_r^3}{\mathcal{E}_r^3} \frac{l_\mathrm{B}^2}{r^2} + 1252\chi^2 \frac{E_r^2}{\mathcal{E}_r^2} \frac{l_\mathrm{B}^4}{r^4} \right)}_{r^{-8}} 
+\, O(r^{-10}) \nonumber\\
&= \omega_\mathrm{c} \left[ 
\frac{2E_r^2}{\mathcal{E}_r^2} \left( 1 + \frac{16\chi l_\mathrm{B}^2}{r^2} + \frac{576\chi l_\mathrm{B}^4}{r^4} \right) 
+ \frac{8E_r^3}{\mathcal{E}_r^3} \left( 1 + \frac{48\chi l_\mathrm{B}^2}{r^2} \right) 
+ \frac{30E_r^4}{\mathcal{E}_r^4}
\right]
+ O(r^{-10}).  
\end{align}
Then, we obtain the azimuthal current density 
\begin{align}
j_\theta &= e \rho v_\theta \nonumber\\
&= e \bar{\rho} \left( 1 + \frac{\omega}{\omega_\mathrm{c}} \right) v_\theta \nonumber\\
&= -\frac{\nu e^2}{h} E_r 
\bigg[
1 
+ \frac{E_r}{\mathcal{E}_r} \left( 1 + \frac{8\chi l_\mathrm{B}^2}{r^2} + \frac{192\chi^2 l_\mathrm{B}^4}{r^4} \right) 
+ \frac{4E_r^2}{\mathcal{E}_r^2} \left( 1 + \frac{24\chi l_\mathrm{B}^2}{r^2} \right)
+ \frac{15E_r^3}{\mathcal{E}_r^3}
\bigg]
+ O(r^{-9}), 
\end{align}
where we use Equation~13. 
We can see that the dimensionless parameter $\chi$ gives the corrections in terms of $\chi (l_\mathrm{B}/r)^2$.  
We now recall the slowly varying condition $l \gg l_\mathrm{B}$.  Here, the characteristic length scale is the radius $r$, and hence we have $r \gg l_\mathrm{B}$.  Therefore, the corrections from $\chi$ are parametrically small, which results in 
\begin{equation}
j_\theta = -\frac{\nu e^2}{h} E_r 
\left(
1 + \frac{E_r}{\mathcal{E}_r} + \frac{4E_r^2}{\mathcal{E}_r^2} + \frac{15E_r^3}{\mathcal{E}_r^3} 
\right)
+ O(r^{-9}).  
\label{eq:S_response-inverse}
\end{equation}

\subsection{Electric field proportional to $\bm{r}$}
\label{sec:S_r}

Next we discuss the flow under the external electric field 
\begin{equation}
\label{eq:S_r}
E_r = a' r.  
\end{equation}
We note that the realization of this electric field is not as simple as the previous example of $E_r \propto 1/r$ because Gauss's law in Maxwell's equations requires a finite charge density $\varepsilon \nabla \cdot \bm{E} = 2a' \varepsilon$ ($\varepsilon$: permittivity) to support this electric field.  Nevertheless, we assume that Equation~\ref{eq:S_r} is realized locally.  
As the electric field is not singular at the origin, the result may be applicable to the flow near the center of a disk, but yet we should remember the condition $r \gg l_\mathrm{B}$.  

An important observation here is that we can exactly solve Equation~\ref{eq:S_hydrodynamic_polar}.  
To show this, we first assume that the azimuthal velocity $v_\theta$ is also proportional to the radius $r$:
\begin{equation}
v_\theta = c' r.  
\end{equation}
Then, the pressure gradient and Hall viscosity contributions vanish 
\begin{equation}
\partial_r \left( \frac{1}{r} \partial_r (r v_\theta) \right) = 0, 
\end{equation}
and the hydrodynamic equation (Equation~\ref{eq:S_hydrodynamic_polar}) becomes 
\begin{equation}
- m \frac{v_\theta^2}{r}= e (E_r + v_\theta B).  
\end{equation}
The dimensionless parameter $\chi$ does not appear in the equation of motion.  
This equation is merely a quadratic equation with respect to $v_\theta$, and its solution is 
\begin{equation}
\label{eq:S_v_polar_analytic}
v_\theta = - \frac{\mathcal{E}_r}{2B} \left( 1 - \sqrt{1 - \frac{4E_r}{\mathcal{E}_r}} \right).  
\end{equation}
We note that $v_\theta$ is indeed proportional to $r$, as both $E_r$ and $\mathcal{E}_r$ are proportional to $r$.

With the velocity obtained, we find the vorticity  
\begin{equation}
\omega = \frac{1}{r} \partial_r (r v_\theta) = \frac{2v_\theta}{r}, 
\end{equation}
which is constant.  Correspondingly, the density $\rho$ also becomes constant: 
\begin{equation}
\rho = \bar{\rho} + \frac{\nu m}{h} \omega = \frac{\nu e B}{h} \sqrt{1 - \frac{4E_r}{\mathcal{E}_r}}.  
\end{equation}
The density distribution is uniform, but its value is different from $\bar{\rho}$, reflecting the constant vorticity of the flow.  
We finally obtain the azimuthal current response 
\begin{align}
j_\theta &= e \rho v_\theta \nonumber\\
&= - \frac{\nu e^2}{h} \frac{\mathcal{E}_r}{2} \left( \sqrt{1 - \frac{4E_r}{\mathcal{E}_r}} - 1 + \frac{4E_r}{\mathcal{E}_r} \right) \nonumber\\
&= - \frac{\nu e^2}{h} E_r \left[ 1 - \frac{E_r}{\mathcal{E}_r} - \frac{2E_r^2}{\mathcal{E}_r^2} - \frac{5E_r^3}{\mathcal{E}_r^3} + O\left( \frac{E_r^4}{\mathcal{E}_r^4} \right) \right].  
\label{eq:S_r_current}
\end{align}
The nonlinear contributions here appear with the opposite sign from Equation~\ref{eq:S_response-inverse}, implying that the nonlinear response relies on the details of the electric field distribution.  

The electric field profile (Equation~\ref{eq:S_r}) has a stronger field for large $r$ and vanishes at $r = 0$.  A nonlinear response in a Hall bar geometry may show a similar behavior as Equation~\ref{eq:S_r_current} rather than Equation~\ref{eq:S_response-inverse}.  The electric field in a device should depend on the details of the geometry and the current flow, and so does the current response.  
For a more realistic and rigorous calculation, we should simultaneously solve the hydrodynamic equations and Maxwell's equations with the boundary conditions that reflect the sample.

\subsection{Discussions}

As we have seen in Sections~\ref{sec:S_1/r} and \ref{sec:S_r}, the effect of the gradient term, namely the term with the coefficient $\chi$ in the equation of motion, is minor for $l \gg l_\mathrm{B}$.  
While both the centrifugal force and the gradient term may induce the corrections to the current response, we have observed that the latter is negligible under the slowly varying conditions Equation~12.  Now we perform an order-of-magnitude estimate of the two corrective contributions and estimate their ratio: 
\begin{equation}
\frac{\hbar \chi \partial_r \left( \dfrac{1}{r} \partial_r (r v_\theta) \right)}{m \dfrac{v_\theta^2}{r}}
\sim \frac{\hbar \chi \dfrac{v_\theta}{l^2}}{m \dfrac{v_\theta^2}{r}}
\sim \chi \left( \frac{l_\mathrm{B}}{l} \right)^2 \frac{\omega_\mathrm{c}}{\omega}.  
\end{equation}
From this estimate, we cannot conclude that the gradient term is always negligible, since $(l_\mathrm{B}/l)^2$ is small but $\omega_\mathrm{c}/\omega$ is large.  

Nevertheless, when we assume that the gradient term is negligible, the equation of motion (Equation~\ref{eq:S_hydrodynamic_polar}) becomes an algebraic equation but not a differential equation, so that we can obtain the analytic expression of the current response.  
It is then legitimate and also instructive to consider the expansion with respect to $E_r/\mathcal{E}_r$, which illustrates the difference between Sections~\ref{sec:S_1/r} and \ref{sec:S_r}.  

As we put $\chi = 0$, the analytic expression of the azimuthal velocity is Equation~\ref{eq:S_v_polar_analytic}, wherein the radial electric field $E_r(r)$ now have arbitrary dependence on $r$.  
By expanding it with respect to $E_r/\mathcal{E}_r$, we obtain 
\begin{align}
v_\theta(r) &= - \frac{E_r}{B} \frac{\mathcal{E}_r}{2E_r} \left( 1 - \sqrt{1 - \frac{4E_r}{\mathcal{E}_r}} \right) \nonumber\\
&= - \frac{E_r}{B} \left[ 1 + \frac{E_r}{\mathcal{E}_r} + \frac{2E_r^2}{\mathcal{E}_r^2} + \frac{5E_r^3}{\mathcal{E}_r^3} + O\left( \frac{E_r^4}{\mathcal{E}_r^4} \right) \right] \quad \text{for any $E_r=E_r(r)$}.  
\end{align}
We then calculate the vorticity following Equation~\ref{eq:S_vorticity_polar}.  Since it contains the derivative with respect to $r$, it depends on the functional form of $E_r$:  
\begin{align}
\omega(r) &= 2\omega_\mathrm{c} \left[ \frac{E_r^2}{\mathcal{E}_r^2} + \frac{2E_r^3}{\mathcal{E}_r^3} + O\left( \frac{E_r^4}{\mathcal{E}_r^4} \right) \right] \hspace{-50pt} & &\text{for $E_r \propto r^{-1}$}, \\
\omega(r) &= -2\omega_\mathrm{c} \left[ \frac{E_r}{\mathcal{E}_r} + \frac{E_r^2}{\mathcal{E}_r^2} + \frac{E_r^3}{\mathcal{E}_r^3} + O\left( \frac{E_r^4}{\mathcal{E}_r^4} \right) \right] \hspace{-50pt} & &\text{for $E_r \propto r$}.  
\end{align}
Therefore, we obtain the different expansions in Equations~\ref{eq:S_response-inverse} and \ref{eq:S_r_current} for the electric fields $E_r \propto r^{-1}$ and $E_r \propto r$, respectively.  The difference arises not from the azimuthal velocity but from the vorticity, which modifies the density distribution.

\end{document}